\documentclass[twocolumn,showpacs,pra,aps,citeautoscript,amsmath,amssymb,floatfix,superscriptaddress]{revtex4-1}
\usepackage[english]{babel}
\usepackage{txfonts}
\usepackage[latin1]{inputenc} 
\usepackage{textcomp}     
\usepackage[T1]{fontenc}
\usepackage{comment}

\usepackage{natbib}

\usepackage{hyperref}
\usepackage{xcolor}
\definecolor{dark-red}{rgb}{0.4,0.15,0.15}
\definecolor{dark-blue}{rgb}{0.15,0.15,0.4}
\definecolor{medium-blue}{rgb}{0,0,0.5}
\hypersetup{
colorlinks, linkcolor={dark-blue},
citecolor={dark-blue}, urlcolor={medium-blue}
}

\usepackage[markup=underlined]{changes}

\definechangesauthor[color=dark-red]{j}

\newcommand{\braket}[1]{\left< #1 \right>}
\renewcommand{\l}{\left(}
\renewcommand{\r}{\right)}
\renewcommand{\b}[1]{\boldsymbol{#1}}

\usepackage{graphicx}
\graphicspath{{pics/}} 
\usepackage{wrapfig}
\usepackage[caption=false]{subfig}
\usepackage{subfloat}
\usepackage{isotope}
\usepackage{dcolumn}
\usepackage{color}
\usepackage{hyperref} 
\usepackage{placeins}
\DeclareGraphicsExtensions{.pdf,.png,.jpg}

\usepackage{empheq} 

\usepackage{mathrsfs} 
\usepackage{amssymb}
\usepackage{amsfonts}
\usepackage{bbm}
\usepackage{bbold}
\usepackage{amsmath}
\usepackage{empheq} 
\usepackage{braket}
\usepackage{mathtools}
\makeatletter
\newcommand{\subalign}[1]{%
  \vcenter{%
    \Let@ \restore@math@cr \default@tag
    \baselineskip\fontdimen10 \scriptfont\tw@
    \advance\baselineskip\fontdimen12 \scriptfont\tw@
    \lineskip\thr@@\fontdimen8 \scriptfont\thr@@
    \lineskiplimit\lineskip
    \ialign{\hfil$\m@th\scriptstyle##$&$\m@th\scriptstyle{}##$\crcr
      #1\crcr
    }%
  }
}
\makeatother

\begin{document}

\title{Dissipative Dicke Model with Collective Atomic Decay: Bistability, Noise-Driven Activation and 
Non-Thermal First Order Superradiance Transition}

\author{Jan Gelhausen}
\email{jg@thp.uni-koeln.de}
\affiliation{Institut f\"ur Theoretische Physik, Universit\"at zu K\"oln, D-50937 Cologne, Germany}

\author{Michael Buchhold}
\email{buchhold@caltech.edu}
\affiliation{Institut f\"ur Theoretische Physik, Universit\"at zu K\"oln, D-50937 Cologne, Germany}
\affiliation{Department of Physics and Institute for Quantum Information and Matter, California Institute of Technology, Pasadena, CA 91125, USA}

\date{\today}

\begin{abstract}
The Dicke model describes the coherent interaction of a laser-driven ensemble of two level atoms with a quantized light field. It is realized within cavity QED experiments, which in addition to the coherent Dicke dynamics feature dissipation due to e.g. atomic spontaneous emission and cavity photon loss. 
Spontaneous emission supports the uncorrelated decay of individual atomic excitations as well as the enhanced, collective decay of an excitation that is shared by $N$ atoms and whose strength is determined by the cavity geometry. We derive a many-body master equation for the dissipative Dicke model including both spontaneous emission channels and analyze its dynamics on the basis of Heisenberg-Langevin and stochastic Bloch equations.
We find that the collective loss channel leads to a region of bistability between the empty and the superradiant state. Transitions between these states are driven by non-thermal, markovian noise. The interplay between dissipative and coherent elements leads to a genuine non-equilibrium dynamics in the bistable regime, which is expressed via a non-conservative force and a multiplicative noise kernel appearing in the stochastic Bloch equations. We present a semiclassical approach, based on stochastic nonlinear optical Bloch equations, which for the infinite-range Dicke Model become exact in the large-$N$-limit. The absence of an effective free energy functional, however, necessitates to include fluctuation corrections with $\mathcal{O}(1/N)$ for finite $N<\infty$ to locate the non-thermal first-order phase transition between the superradiant and the empty cavity.
\end{abstract}

\maketitle
\section{Introduction}
Bistable interacting quantum systems are prime examples, for which fluctuation induced many-body effects beyond the mean-field picture dominate the long-time dynamics. Quantum fluctuations induce rare transitions between the mean-field steady states, rendering them metastable and introducing macroscopic effects beyond small perturbations \cite{Abbaspour15,Kerckhoff11}.

Adding the concept of drive and dissipation introduces an additional competition between unitary and dissipative dynamics, which has a pronounced impact on the fluctuation induced, asymptotic dynamics. This was early observed for a driven cavity with an optical Kerr-nonlinearity \cite{Bonifacio1982}. On the level of single-operator expectation values the system shows a coexistence regime of two stable states with different photon numbers for the same driving strength. However, a full quantum treatment \cite{Drummond1980,Kheruntsyan1999} reveals that quantum fluctuations induce a driven-dissipative first-order phase transition in the thermodynamic limit of large intracavity photon numbers \cite{Casteels17},  recently observed in a semiconductor microcavity \cite{rodriguez2017}.

This growing experimental access and realization of driven-dissipative non-linear systems out of equilibrium that display bistable dynamics \cite{Rempe,Baas04,Hruby17}, including recently detected bistability regions in cavity QED \cite{Angerer171} and circuit QED \cite{Fink17} experiments, has led to a surge of interest in theoretical descriptions for out of equilibrium systems \cite{Buchhold17,sieberer2016,Hafezi17,Buchhold13,Torre13,Maghrebi16}. This includes signatures of first-order dissipative phase transitions \cite{Fink20172,Casteels16,Kessler12,Carmichael15} and the crucial role played by fluctuations \cite{Marcuzzi16}.

Here, we extend the simple, yet paradigmatic Dicke model \cite{Dicke1954}, describing light-matter interactions of $N$-atoms coupled to a single quantized photon mode, to the dissipative regime. By considering both dissipative single-particle and cooperative effects, we devise a quite generic extension of the model, which respects {\it(i)} the $\mathbb{Z}_2$-Ising symmetry, {\it (ii)} the steady state manifold of a superradiant and a dark atomic ensemble and {\it (iii)} locality in time. This generic and in this sense universal modification opens up a novel dynamical regime, which we demonstrate to be dominated by macroscopic atomic fluctuations and the absence of detailed balance.

In the context of cavity QED, the dissipative Dicke model emerges when single atom and collective decay of excitations into the electromagnetic vacuum are considered beyond a single-excitation framework. It extends recent works on super- and subradiant cavity states \cite{Scully2015}, and the resonance fluorescence model \cite{Drummond78} to the experimentally relevant many-body regime with strong atom-light coupling \cite{Angerer171,Zhiqiang17}.

The driven-dissipative Dicke model is through its simplicity exceptionally well-suited to study universal non-equilibrium behavior close to a first-order phase transition. Through the presence of a large number of cavity-emitters, the large-$N$ limit is well-controlled theoretically for both mean-field and fluctuation dynamics. The model is therefore a prime candidate to study non-thermal noise-activation, the fate of hysteresis and bistability for experimentally relevant conditions. 

If the cooperative dissipation exceeds a critical value, we find that the conventional second-order phase transition from the empty to the superradiant state is replaced by a bistable regime for the dark and the bright cavity, which is absent for weak losses \cite{Gelhausen17}. 
Mapping the quantum dynamics to semiclassical stochastic optical Bloch equations, we show that the bistable regime features noise activated transitions between the metastable states.
The activation rates are, however, suppressed exponentially with the number of atoms, such that in the thermodynamic limit one of the metastable states becomes stable and thus the true steady state. This leads to a collapse of the bistable regime towards a sharp first-order transition. Due to the absence of detailed balance the dynamics in this regime can neither be derived from the gradient of a potential, as is the case in dissipative equilibrium, nor are fluctuations uniformly distributed in phase space but strongly state-dependent. We give a brief discussion of the difference between effective equilibrium and non-equilibrium systems on the basis of the Martin-Siggia-Rose-Janssen-de Dominicis (MSRJD) path integral framework below.

\section{Quantum Master Equation}The state of the cavity ensemble of $N$ atoms and a single photon mode is expressed via the density matrix $\rho$, whose time evolution is given by the Markovian quantum master equation $\partial_t\rho=-i[H,\rho]+\mathcal{L}_\gamma[\rho]+\mathcal{L}_\kappa[\rho]$.
The coherent evolution, including the cavity photon-atom coupling, is given by the Dicke Hamiltonian \cite{Dicke1954,garraway2011}
\begin{align}
H&=\omega_0 a^{\dagger}a
+\frac{\omega_z}{2}\sum_{\ell=1}^{N}\sigma^z_{\ell} +\frac{g}{\sqrt{N}}(a+a^{\dagger})\sum_{\ell=1}^{N}\sigma^{x}_{\ell}\;.
\label{EQ:Hamiltonian}
\end{align}
Here, $\omega_0, \omega_z$ is the photon energy and the atomic level splitting and $g$ is the atom-light coupling strength. Equation \eqref{EQ:Hamiltonian} implicitly contains an external drive laser, whose time-dependence has been eliminated in a rotating frame \cite{Dimer07}.
\begin{figure}
\includegraphics[width=8.5cm]{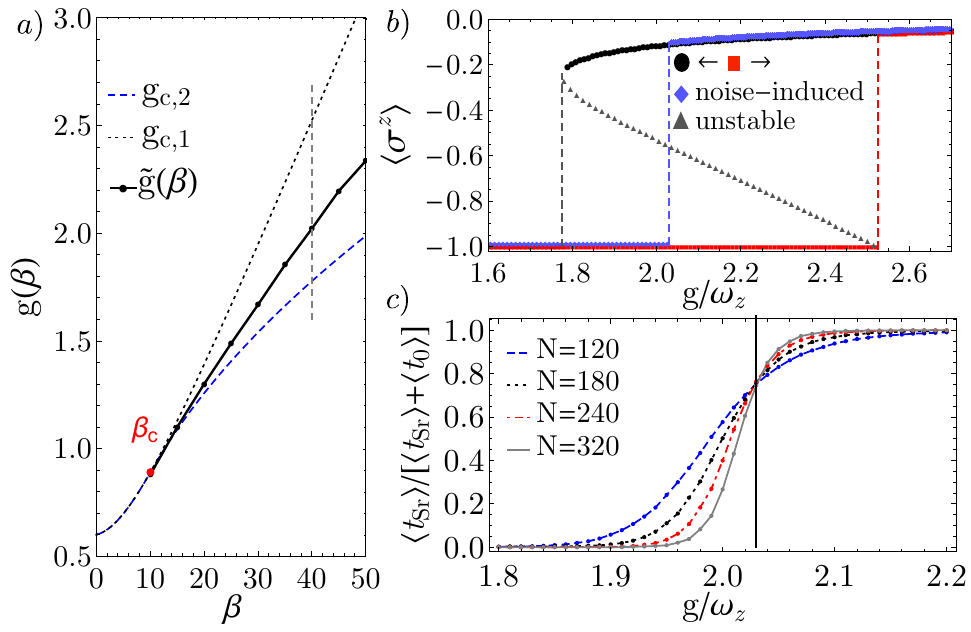}
\caption{
a) Mean-field bistability region delimited by the spinodal (broken) lines and fluctuation induced first order transition (bold) line at  $\tilde{g}(\beta)$. Vertical line cuts through the phase diagram as shown in b) and c). b) (Mean-field) Hysteresis inside the bistable regime obtained by adiabatically following the increasing (red), decreasing (black) atom-light coupling. c) Average amount of time spent in the superradiant state obtained from stochastic Bloch equations for different atom numbers $N$, see Eq.\,\eqref{EQ:LangevinEquations}. For $N \to \infty$ it approaches a step function, revealing a first-order transition at $\tilde{g}$ (intersection), as shown in a).  Parameter set $\omega_0=1.4\omega_z$, $\kappa=2\gamma=0.2\omega_z$, with $\beta=40$.
}
\label{Fig:Spinodallines}
\end{figure}

The decay into the weakly coupled and thermodynamically large photon vacuum are well treated in the Born-Markov approximation (see App.\,\ref{App:SecCollectiveAtomicDecay}), leading to the time-local Lindblad
\begin{align}
\mathcal{L}_\gamma[\rho]=&\gamma(1-\beta/N)\sum_{\ell=1}^N \left( 2\sigma^{-}_{\ell}\rho\sigma^{+}_{\ell}-\{\sigma^{+}_{\ell}\sigma^{-}_{\ell},\rho\}\right) \label{EQ:AtomicLindbladOperator}\\
&+\gamma \beta/N \big( 2S^{-}\rho S^{+}-\{S^{+}S^{-},\rho\}\big),\label{eq3} \\
\mathcal{L}_\kappa[\rho]&=\kappa\big(2 a \rho a^{\dagger}-\{a^{\dagger}a ,\rho\}\big)\label{EQ:PhotonicLindbladOperator}\;.
\end{align}
The two-level atoms are modeled by a local spin algebra $\sigma^+_i=\ket{e}_i\bra{g},\sigma^z_i=\ket{e}_i\bra{e}-\ket{g}_i\bra{g}$ and $S^{\pm}=\sum_{\ell=1}^N \sigma^{\pm}_{\ell}$. Here $(e,g)$ refers to the excited and ground state of an atom, respectively and $a^\dagger, a$ creates, annihilates a cavity photon.
Photon loss through the cavity mirrors with rate $\kappa$ is described by the Lindblad $\mathcal{L_{\kappa}}$. Atomic spontaneous emission into the electromagnetic vacuum outside the cavity is captured by $\mathcal{L}_\gamma$.

The atoms can either decay individually and uncorrelated \eqref{EQ:AtomicLindbladOperator} or can decay through a collective channel, resulting from the electromagnetic bath being commonly shared by all the atoms. In the context of the Dicke model, the most important collective decay channel is the spontaneous decay of a superradiant state \eqref{eq3}.
Intuitively, the photon rate of dissipation $\kappa$ and the single atom loss rate $\gamma$ shift the critical atom-light coupling for the superradiant phase transition towards higher pump strength to compensate for the losses  \cite{Gelhausen17,Dimer07,Torre16,Keeling17}. The collective noise, however, introduces an additional non-linearity, leading to a drastic modification of the phase transition in the thermodynamic limit $N\rightarrow\infty$.

When the atomic emitters radiate collectively, the decay rate is actually enhanced by the number of emitters $\gamma \to \gamma N$. This can occur even for a single photonic excitation that is shared among $N$ atoms and is known as single-photon superradiance \cite{Scully09SUperofSuperradiance}. Collective emission of radiation is relevant for many-body states such as atomic, collective angular momentum or Dicke states where the atomic ensemble can behave like one giant atom. Geometrically, this behaviour is typically expected when the atoms are closer together than the wavelength of radiation. However, even for larger atomic samples the radiation rate can be enhanced \cite{Scully06,Scully2015}. In the present case the relevant geometric factor $(\alpha\ll1)$ depends on the size of the atomic cloud and the cavity parameters, see  App.\,\ref{App:SecCollectiveAtomicDecay}. The collective excitation rate is then modified as $\gamma N \to \gamma \alpha N\equiv \gamma \beta$. However, both the average energy and loss rate per particle have to remain finite in the thermodynamic limit, which implies that for $N \to \infty$, $\beta=const.$ and is set by the fixed number of atoms in an experiment. Both the collective and the individual loss channel are derived from the same Hamiltonian that couples the system degrees of freedom with the electromagnetic vacuum, see App.\,\ref{App:SecCollectiveAtomicDecay}. The collective atomic loss channel does therefore not introduce any new characteristic time scales that would call the time-local Lindblad structure and thus the Born-Markov approximations into question that are valid when system-bath couplings are small $\gamma\ll (\omega_0,\omega_z,g)$.


\section{Heisenberg-Langevin analysis}
In order to derive the semiclassical optical Bloch equations for the dissipative Dicke model (Eqs.~\eqref{EQ:Hamiltonian}-\eqref{EQ:PhotonicLindbladOperator}), which can be addressed numerically, we start with the computation of the Heisenberg-Langevin  equations (HLE) for the individual spin and photon operators.
Although the Dicke Hamiltonian in Eq.\,\eqref{EQ:Hamiltonian} preserves the total spin quantum number of the system, the single atom loss components mix sectors of different total spin $\hat{S}^2$ and the HLE have to be expressed in the local spin basis. 

\label{subsec:langevin}
The HLE are obtained from a conjugate master equation
\begin{align}
\partial_t O_i=-i[O_i,H]+
\mathcal{L}^*_\gamma[O_i]+\mathcal{L}^*_\kappa[O_i]+\xi^O_{i},\label{eq5}
\end{align}
where in $\mathcal{L}^*$ compared to $\mathcal{L}$ the order of operators in the quantum jump term has been inverted, e.g. $\mathcal{L}^*_\kappa[.]=\kappa(2a^\dagger .\ a-\{a^\dagger a, . \})$. The quantum noise operators $\xi^O_i$ ensure the correct time evolution of fluctuations $O_iO_l$ and preserve the exact operator commutation relations \cite{Scully97}, for the explicit derivation see App.\,\ref{App:ClassicalNoiseKernel}.
Applying Eq.~\eqref{eq5} to the photon field and the individual spin components at site $i$ yields
\begin{align}
\partial_t a&=-\l \kappa+i\omega_0\r a- \frac{ig}{\sqrt{N}} \sum_{\ell=1}^N \sigma^x_{\ell}+\xi^a,\label{EQ:HeisenbergLangevinPhoton}\\
\partial_t \sigma^x_i&=-\omega_z\sigma^y_i -\gamma\sigma^{x}_i+\frac{\gamma\beta}{N}\sum_{\ell\neq i} \sigma^{x}_{\ell}\sigma^{z}_{i}+\xi^x_i,\\
\partial_t \sigma^y_i&=\omega_z\sigma^x_i -\left[\frac{2g(a^\dagger+a)}{\sqrt{N}}-\frac{\gamma \beta}{N}\hspace{-0.1cm}\sum_{\ell\neq i} \sigma^{y}_{\ell}\right]\sigma^z_i-\gamma\sigma^{y}_i+\xi^y_i, \\
\partial_t \sigma^z_i&=\frac{2g(a^{\dagger}+a)}{\sqrt{N}}\sigma^y_i-2\gamma(1+\sigma^z_i)+\xi^z_i \label{EQ:HeisenbergLangevinPhotonSigmaz}\\
&-\frac{\gamma\beta}{2N} \sum_{\ell\neq i}\l \sigma^{x}_{i}\sigma^{x}_{\ell}+\sigma^{y}_{i}\sigma^{y}_{\ell}+i(\sigma^y_{i}\sigma^{x}_{\ell}-\sigma^{x}_{i}\sigma^{y}_{\ell})+cc. \r. \nonumber
\end{align} 
Eliminating the gapped photon field by setting ${\partial_ta=0}$ in \eqref{EQ:HeisenbergLangevinPhoton} and solving for $a$ adds a nonlinear, ferromagnetic coupling $\sim -J/N \sum_\ell \sigma^x_\ell$, with ${J=4g^2\omega_0/(\kappa^2+\omega_0^2)}$, and the photon noise $\xi^a+\xi^{a^\dagger}$ to the atomic HLE. %
The collective decay leads to another, competing nonlinearity $\sim \gamma\beta$ to the HLE, which introduces a bistable regime for parameters $\gamma \beta \approx J \approx J_c$, where $J_c$ is the critical coupling for the superradiance transition. We analyze the remaining set of equations for the atoms in a large $N$-framework, which has been shown to be in good agreement with experimental measurements, see e.g.\,\cite{Angerer171,Zhiqiang17,baumann2010dicke,Esslinger17}. 
The operators in the HLE are replaced by the quantum mechanical average over all atoms, i.e. we analyze the equations of motion for $\sigma^\alpha=\sum_\ell \langle\sigma^\alpha_\ell\rangle/N$, $\alpha=x,y,z$. Approximating the average of the double sums $\sum_{i,\ell}\langle\sigma^\alpha_i\sigma^\beta_\ell\rangle/N^2$ by the product $\sigma^\alpha\sigma^\beta$ is correct up to $\mathcal{O}(1/N)$ corrections and becomes exact in the thermodynamic limit. This is due to the infinite range of both the Dicke nonlinearity $\sim g$ and the collective loss $\sim \gamma\beta$.

Disregarding the noise yields the deterministic optical Bloch equations $
\partial_t\sigma^{\alpha}=D^{\alpha}$ 
with the deterministic force 
\begin{align}\label{EQ:DeterministicDrift}
\b{D}=\left(
\begin{array}{c}
 -\gamma  \sigma^x (1-\beta  \sigma^z)-\sigma^y \omega_z \\
 \omega_z \sigma^x+J \sigma^x \sigma^z- \gamma  \sigma^y (1-\beta  \sigma^z) \\
 -2\gamma(\sigma^z+1)-\gamma \beta  \left((\sigma^x)^2+(\sigma^y)^2\right)-J\sigma^x \sigma^y \\
\end{array}
\right),
\end{align}
including the additional nonlinearities $\sim J,\beta \gamma$ in comparison to the conventional Bloch equations \cite{Scully97}.

We stress here that $\nabla_{\sigma} \times \b{D}\neq 0$ and $\nabla_\sigma \cdot \b{D}\neq 0$, which results from the presence of unitary and dissipative dynamics and prohibits the interpretation of $\b{D}$ as a conservative force, $\b{D}\neq\nabla_\sigma V$ for some potential $V$. Here, unitary and dissipative dynamics cannot be generated by the same Hamiltonian, which excludes a dissipative equilibrium, where steady states coincide with minimums in a generalized energy landscape, see Sec.\,\ref{Sec:NonEqMSRJD}. 

Solving $\b{D}=0$ yields the mean-field stationary states and determines the steady state values for the $\sigma^{\alpha}$. For the population imbalance, one finds (considering only real solutions)
\begin{align}{\label{EQ:SigmazNoiselessDynamics}}
\sigma^z\hspace{-0.05cm}=\hspace{-0.05cm}\max\Big\{\hspace{-0.05cm}-\hspace{-0.05cm}1,\tfrac{J\omega_z}{2\beta^2\gamma^2}\big[(1\hspace{-0.05cm}-\hspace{-0.05cm}4\beta \gamma^2/J\omega_z\hspace{-0.05cm}-\hspace{-0.05cm}4\beta^2\gamma^2/J^2)^{\frac{1}{2}}\hspace{-0.05cm}\hspace{-0.05cm}-\hspace{-0.05cm}1\big]\hspace{-0.05cm}+\hspace{-0.05cm}\tfrac{1}{\beta}\Big\}.
\end{align}
For collective loss strengths $\beta<\beta_c=\sqrt{1+(\omega_z/\gamma)^2}$, $\sigma^z$ is a continuous function of the Dicke coupling $g$. For coupling strengths above a critical value $g\ge g_{c,1}$ (Eq.~\eqref{EQ:criticalcoupling1}), $\sigma^z\sim |g-g_{c,1}|^{\nu_x}-1$ deviates from its empty cavity value of $\sigma^z=-1$. This goes in hand with a macroscopic occupation of the cavity mode $\langle a^{\dagger}a\rangle\sim N|g-g_{c,1}|^{\nu_x}$, i.e. the continuous phase transition from the empty cavity to the superradiant state. Here $\nu_x=1$ is the finite temperature photon flux critical exponent \cite{Torre13}. 

Above the critical loss strength, $\beta>\beta_c$, the continuous transition into the superradiant state is replaced by a discontinuous jump of $\sigma^z$ at $g=g_{c,1}$ with magnitude $\propto 1-\beta_c/\beta$. A closer look at the optical Bloch equations reveals that for $g_{c,2}\le g\le g_{c,1}$ (Eq.~\eqref{EQ:criticalcoupling1}) both the empty as well as the superradiant state appear as attractive stationary mean-field solutions. This indicates classical bistability, where steady states are sensitive towards the initial configuration leading to the appearance of hysteresis, Fig.\,\ref{Fig:Spinodallines}b). The critical couplings are
\begin{align}
g_{c,1}&=\gamma\left[\frac{(\kappa^2+\omega_0^2)(\beta_c^2+\beta^2+2\beta) }{4\omega_z  \omega_0}\right]^{\frac{1}{2}}, \ \ \frac{g_{c,2}}{g_{c,1}}=\left[\frac{2\beta(1+\beta_c)}{\beta^2_c+\beta^2+2\beta}\right]^{\frac{1}{2}}\label{EQ:criticalcoupling1}.
\end{align}

\begin{figure}
\includegraphics[width=8cm]{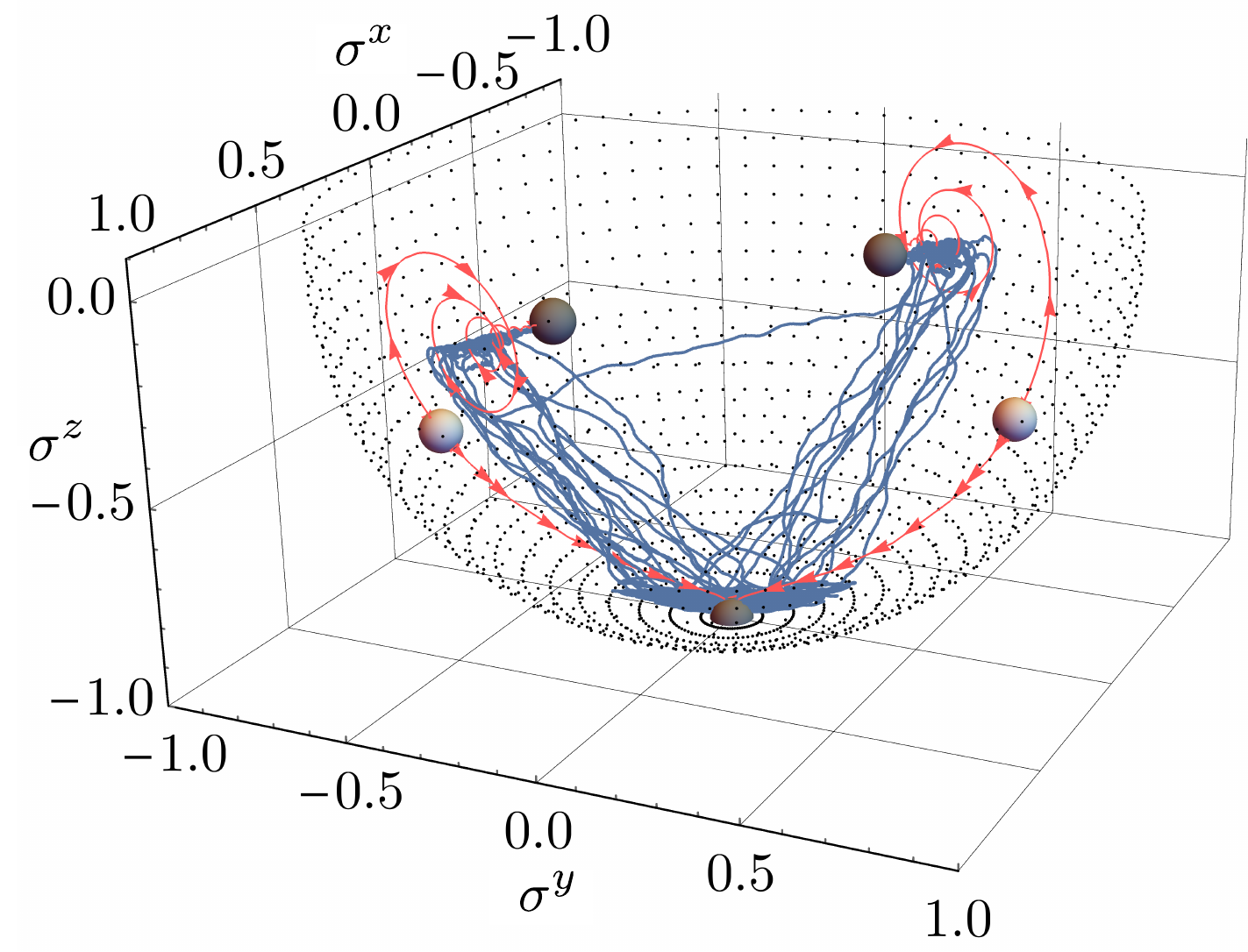}
\caption{
Steady state landscape of the Bloch equations \eqref{EQ:LangevinEquations} in the bistable regime. Red lines (with arrows) departing from repulsive fixed points show deterministic motion, blue lines (no arrows) show noise-induced dynamics. There are 3 attractive (black sphere) and 2 repulsive (gray sphere) fixed points, interrelated by the Ising symmetry $(\sigma^x,\sigma^y,\sigma^z) \to (-\sigma^x,-\sigma^y,\sigma^z)$. Dotted points map out the surface of the unit Bloch sphere and serve as a guide to the eye.
}
\label{fig:stabilitylandscape}
\end{figure}

The bistable regime is delimited by the spinodal lines ($g_{c,1}(\beta),g_{c,2}(\beta)$) that continuously connect at the bicritical point $\beta_c$ as shown in Fig.\,\ref{Fig:Spinodallines}a). Inside this regime, the steady states, corresponding to the empty and the superradiant cavity, appear as attractive fixed points of $\b{D}$. A two-dimensional separatrix divides the Bloch sphere into the two corresponding basins of attraction and hosts a further, unstable fixed point, as illustrated in Fig.\,\ref{fig:stabilitylandscape}. At a spinodal line, the separatrix encloses one of the attractive fixed points and forces it to vanish.

Exactly at the bicritical point, $\beta=\beta_c$, the superradiance transition remains continuous but the critical exponents are different compared to $\beta<\beta_c$, which defines a different universality class for the bictritical point. The photon flux exponent for instance can be inferred from Eq.~\eqref{EQ:SigmazNoiselessDynamics} and reads $\nu_x=\frac{1}{2}$ as for the quantum phase transition in the coherent Dicke model \cite{Nagy11}. A full classification of the bicritical point is left to future work. 


\section{Stochastic Bloch Equations}
Within the bistable regime, the deterministic optical Bloch equations split the Bloch sphere into two basins of attraction, separated by a repulsive manifold, which cannot be crossed by any deterministic path. They thus fail to capture the dynamics of the initial quantum master equation even on the qualitative level since the latter is predicted to have a unique stationary state for any finite $N$. 

This separation, and the enforced lack of ergodicity, is overcome by translating the quantum noise terms in the HLE to a classical noise, which adds to the deterministic part and yields stochastic optical Bloch equations
\begin{align}{\label{EQ:LangevinEquations}}
\partial_t\sigma^{\alpha}=D^{\alpha}+ \xi^{\alpha} N^{-\frac{1}{2}}.
\end{align}
According to the definition $\sigma^{\alpha}=\sum_\ell \langle \sigma^{\alpha}_\ell\rangle/N$, the classical noise $\xi^\alpha=\sum_\ell\langle \xi^{\alpha}_\ell\rangle/\sqrt{N}$. This average must be taken with care. It indicates only the quantum mechanical average of system operators and not the noise average, which corresponds to the expectation value of bath operators. The first and second moments of the noise are $\langle \xi^{\alpha}\rangle_\text{noise}=0$ and ($\langle . \rangle_\text{noise}$ indicating noise average, $\langle . \rangle_{\text{sys+bath}}$ average of system and bath operators)
\begin{align}{\label{EQ:CollectiveNoiseKernel}}
\langle\xi^{\alpha}\xi^{\beta}\rangle_{\text{noise}}=N^{-1}\langle\sum_{\ell,m}\xi^{\alpha}_{\ell}\xi^{\beta}_{m}\rangle_{\text{sys+bath}}=\delta(t-t')\chi^{\alpha \beta}(\b{\sigma}).
\end{align}
In the large-$N$ limit, the covariance matrix is (see App.\,\ref{App:ClassicalNoiseKernel})
\begin{align}
\b{\chi}=2\left(
\begin{array}{ccc}
  \gamma  \left(\beta  (\sigma^z)^2+1\right) & \dots & \dots  \\
 0 & \left(\tilde{\kappa}+\gamma \beta\right) (\sigma^z)^2+\gamma & \dots \\
  \sigma^x (1-\beta  \sigma^z) \gamma  & \sigma^y \left(\gamma-\left(\tilde{\kappa}+\gamma \beta\right) \sigma^z\right) & \chi_{3,3}(\sigma) \\
\end{array}
\right),
\label{EQ:Noise-Kernelexplicit}
\end{align}
where ${\chi_{3,3}(\sigma)=\tilde{\kappa} (\sigma^y)^2+ \gamma  [\beta  ((\sigma^x)^2+(\sigma^y)^2)+2 (\sigma^z+1)]}$, $\tilde{\kappa}=\kappa J/\omega_0$ and  $\b{\chi}=\b{\chi}^{\rm T}$ is symmetric, real and positive semi-definite for $\sum_{\alpha}(\sigma^{\alpha})^2\leq 1$, i.e. as long as $\b{\sigma}$ represents a state within the Bloch sphere. For a continuous time evolution, neither the deterministic force $\b{D}$ nor the noise drives the system out of the Bloch sphere. The latter is ensured by $\b{\chi}_{\perp}\sim ||\delta\b{\sigma}||$, where $\b{\chi}_\perp$ is the (local) perpendicular noise strength and $\delta\sigma$ the distance to the Bloch sphere. 
\begin{figure}[t]
\includegraphics[width=8cm]{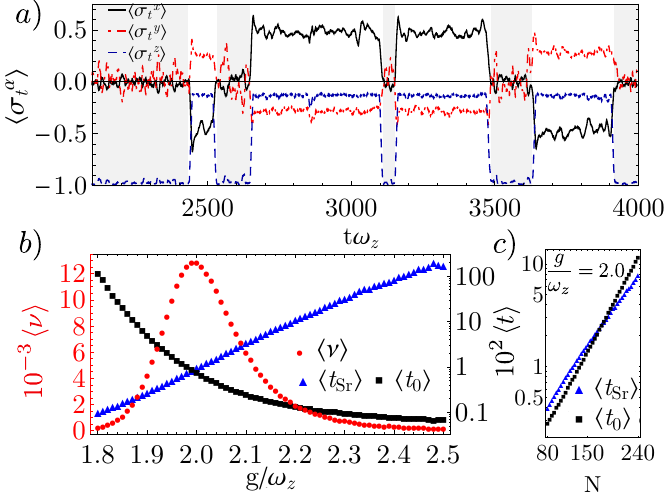}
\caption{(a) Noise-induced transitions from the empty (shaded) to the superradiant state. Each line is a moving average over a time window $(t\pm 4)\omega_z$. (b) Log-plot of mean times spent in the superradiant- $\braket{t_{Sr}}$ and in the empty cavity state $\braket{t_{0}}$ corresponding to vertical line in Fig.\,\ref{Fig:Spinodallines}a). Circular shapes are mean transition frequencies $\braket{\nu}=\#\text{jumps}/t_{\Sigma}$ obtained from counting the total number of jumps normalised to the total simulation time $t_{\Sigma}$  (c) Exponential sensitivity of mean occupation times to the number of atoms $N$. 
}
\label{Fig:NoiseInducedJumps}
\end{figure}
%
For the long-range interacting Dicke system the local noise terms do not break translational invariance and the description in terms of the collective variable $\sigma^{\alpha}$ is correct up to $\mathcal{O}(1/N)$. Locally induced, noise-driven spin flips cause energetic corrections of order $\mathcal{O}(J/N)$ such that for any finite $N<\infty$ the two fixed points can only be connected via a concatenation of $\mathcal{O}(N)$ subsequent noise kicks. Such collective events of noise kicks drive the system from the basin of attraction of one fixed point to the basin of attraction of the other. They occur on timescales set by $\tau N$ where $\tau^{-1}$ is the state-dependent rate of a single spin flip set by the noise profile $\b{\chi}(\sigma)$.

\section{Real-time dynamics of the Bloch equations}
In contrast to a bistable system in equilibrium, where the occupation of the states in the long-time and large system-size limit is entirely determined by a mean-field analysis of the minimum of a free energy potential, the occupation times of the metastable states out-of equilibrium can only be determined beyond mean-field by including fluctuations of $\mathcal{O}(1/N)$ in a numerical simulation for finite $N<\infty$ \cite{Freidlin1984}.
Simulation of Eqs.~\eqref{EQ:LangevinEquations} requires a careful implementation of the thermodynamic limit. Taking $N\rightarrow\infty$ first, leads the noise $\sim 1/\sqrt{N}$ to vanish and one ends up with the deterministic equations, i.e. two disconnected steady states. On the other hand, taking $t\to \infty$ first and then $N\to \infty$, for any finite $N$ the long-time behavior is characterized by an admixture of the empty and the superradiant state. The dynamics of the spin vector $\b{\sigma}$ is obtained by interpreting Eq.\,\eqref{EQ:LangevinEquations} in an It$\hat{o}$-sense and numerically simulating the time-evolution with a two-stage stochastic Runge-Kutta algorithm \cite{Roessler09,wiktorsson2001}. The corresponding dynamics of $\b{\sigma}$ including rare fluctuations between the dark and the bright cavity are visualized in Fig.~\ref{fig:stabilitylandscape} and, with temporal resolution, in Fig.~\ref{Fig:NoiseInducedJumps}a. Similar fluctuation induced switching dynamics have been measured experimentally in bistable semiconductor microcavities \cite{rodriguez2017}.

Tuning the atom-light coupling $g$ through the bistable regime at fixed collective loss rate $\gamma\beta$,  Fig.~\ref{Fig:Spinodallines}a, we obtain a histogram for the distribution of time intervals spent in the empty $(t_{0})$ and in the superradiant state $(t_{Sr})$. The mean occupation times $\braket{t_{\alpha}}$ in Fig.~\ref{Fig:NoiseInducedJumps}b) are obtained by summing over all intervals $\braket{t_{\alpha}}=\sum_{i}t_{i,\alpha}/\sum_{i}$ with $\alpha=\{a, Sr\}$ as seen in Fig.\,\ref{Fig:NoiseInducedJumps}a). We observe an exponential dependence of $\log\braket{t_{Sr}}\sim g$ in the superradiant state and a strongly stretched exponential $\log\braket{t_{0}}\sim g^{-10}$ \textcolor{dark-red}in the empty state, Fig.\,\ref{Fig:NoiseInducedJumps}b. 

For any $N<\infty$, the steady state is a statistical mixture of the empty and the superradiant state and the degree of mixing is expressed via the superradiance occupation ratio $\rho_{Sr}=\braket{t_{Sr}}/(\braket{t_{Sr}}+\braket{t_{0}})$. It interpolates continuously between $\rho_{Sr}(g_{c,2})=0$ and $\rho_{Sr}(g_{c,1})=1$ as a function of the atom-light coupling $g$ and varies on a scale $\Delta g\sim 1/N$, see Fig.~\ref{Fig:Spinodallines}c. In the thermodynamic limit $N\rightarrow\infty$, $\rho_{Sr}\rightarrow \Theta(g-\tilde{g})$ approaches a step-function, indicating a discontinuous jump and a first order phase transition from the empty to the superradiant state at a critical coupling $\tilde{g}(\beta)$, Fig~\ref{Fig:Spinodallines}a. 
The exponential increase of the occupation times $\log \langle t_{Sr,0}\rangle\sim N$ is depicted in Fig.~\ref{Fig:NoiseInducedJumps}c. This suggests a typical Arrhenius law $\langle t_{Sr,0}\rangle= A_{Sr,0}\exp(N\psi_{Sr,0})$, which is confirmed by the classical action \cite{MSR1973,Janssen1976,Dominicis1978} associated to the stochastic optical Bloch equations \eqref{EQ:LangevinEquations}, see Sec.\,\ref{Sec:NonEqMSRJD}. Here, $\psi_\alpha$ is the non-equilibrium potential, which depends both on the path and on the relative noise strength between the two stable solutions and lacks the interpretation of a free energy functional \cite{Freidlin1984}. The Ising symmetry reduces the long time dynamics to that of an effective two-level system, which always fulfills detailed balance, and makes the non-equilibrium nature of the bistability hardly observable on the level of the transition rates. Instead of minimizing an effective free energy \cite{Eyring1935,Landauer1961,Langer1969}, however, the escape trajectories follow the most probable path from one stable fixed point to the other and do not pass the repulsive fixed point, in contrast to equilibrium.  

\section{Non-Equilibrium Apects for  Noise Activated Trajectories in the MSRJD - Path Integral}
\label{Sec:NonEqMSRJD}
In this section, we want to confirm, without being exhaustive, three statements that we made in the previous sections on the non-equilibrium nature of the bistability. The analysis is based on Freidlin-Wentzell theory  \cite{Freidlin1984} for weak noise systems, therefore applicable in the limit $N\rightarrow\infty$, and performed in a Martin-Siggia-Rose-Janssen-de Dominicis (MSRJD) path-integral framework \cite{MSR1973,Janssen1976,Dominicis1978}. Additionally we comment on the structure of the noise-induced trajectories as observed in Fig.\,(2).

\textit{First}, the mean occupation times $\braket{t_0}$ and $\braket{t_{Sr}}$ of the empty and the superradiant state obey an Arrhenius law of the type $\braket{t_{Sr,0}}\sim A_{Sr,0}\exp(N\psi_{Sr,0})$, with $\psi_{Sr,0}(\sigma)$ as a non-equilibrium potential that measures the cost of fluctuations. 

\textit{Second},
The deterministic force $\b{D}$, Eq.~(10), and the noise kernel $\b{\chi}$, Eq.~(17), do not satisfy the necessary condition for microscopic reversibility and violate detailed balance.

\textit{Third}, we comment that the combination of $(\nabla \times D \neq 0)$ and $(\nabla \cdot D \neq 0)$ is a necessary but not a sufficient criterion for out-of-equilibrium dynamics, which do not relax towards an effective thermal equilibrium. 

We start with the MSRJD action, see e.g. \cite{MSR1973,Janssen1976,Dominicis1978}, associated to the stochastic optical Bloch equations (15) ($\alpha=x,y,z$)
\begin{align}\label{actN}
S&=N\int_t  \left[\tilde{\sigma}^\alpha\partial_t\sigma^\alpha-\tilde{\sigma}^\alpha D^\alpha-\frac{1}{2}\tilde{\sigma}^\alpha\chi^{\alpha\beta}\tilde{\sigma}^\beta\right]\\
&=N\int_t  \left[\tilde{\sigma}^\alpha\partial_t\sigma^\alpha-\mathcal{H}\right], \ \ \ Z=\int\mathcal{D}[\{\sigma^\alpha,\tilde{\sigma}^\alpha\}]\ e^{-S}
\end{align}
Here, $D$ and  $\chi$ are given by Eqs.\,(10), (17) and 
\begin{align}
\mathcal{H}=\tilde{\sigma}^\alpha D^\alpha+\frac{1}{2}\tilde{\sigma}^\alpha\chi^{\alpha\beta}\tilde{\sigma}^\beta
\end{align}
is the Freidlin-Wentzell Hamiltonian, $Z$ is the partition function and the fields $\tilde{\sigma}^\alpha$ are the so-called MSRJD response fields. In the limit of $N\rightarrow\infty$ only the saddle points of $S$ contribute to $Z$. The associated equations of motion demonstrate that  $(\tilde{\sigma}^{\alpha},\sigma^{\alpha})$ are canonically conjugate variables
\begin{align}\label{Saddledyn}
0&=\frac{1}{N}\frac{\delta S}{\delta\tilde{\sigma}^\alpha}=\partial_t\sigma^\alpha-\frac{\partial\mathcal{H}}{\partial\tilde{\sigma}^\alpha},\\
0&=\frac{1}{N}\frac{\delta S}{\delta\sigma^\alpha}=-\partial_t\tilde{\sigma}^\alpha-\frac{\partial\mathcal{H}}{\partial\sigma^\alpha},
\end{align}
showing that the Hamiltonian itself is an integral of motion, $\partial_t \mathcal{H}=0$. All saddle-point trajectories starting from a deterministic field configuration, $\tilde{\sigma}^\alpha=0$, fulfill $\mathcal{H}=0$. The explicit form of Eqs.~\eqref{Saddledyn} is
\begin{align}
\partial_t \sigma^{\alpha}=D^\alpha+\chi^{\alpha \beta}\tilde{\sigma}_{\beta}, \ \ \ \partial_t \tilde{\sigma}^{\alpha}&=-\tilde{\sigma}_{\beta}\frac{\delta D_{\beta}}{\delta \sigma^{\alpha}}-\frac{1}{2}\tilde{\sigma}_{\gamma}\left(\frac{\delta\chi_{\gamma \delta}}{\delta \sigma^{\alpha}}\right)\tilde{\sigma}_{\delta}.\label{EQ:DeterministicSaddlePoint2}
\end{align}
The solution of $\mathcal{H}=0$ with $\tilde{\sigma}^{\alpha}=0$, recovers indeed the noiseless, deterministic Bloch equations, $\partial_t \sigma^{\alpha}=D^\alpha$ and $\partial_t\tilde{\sigma}^\alpha=0$.

Noise activated trajectories, in turn, correspond to solutions with $\tilde{\sigma}^\alpha\neq0$. After some algebra one finds that the second equation in \eqref{EQ:DeterministicSaddlePoint2} requires $\b{\tilde{\sigma}}$ to be of the form $\b{\tilde{\sigma}}=\nabla_{\sigma} \Gamma(\b{\sigma})$. The scalar potential $\Gamma$ is defined by the Hamilton-Jacobi equation
\begin{align}
\mathcal{H}(\sigma,\nabla \Gamma)=\partial_{\alpha}\Gamma(D^{\alpha}+\frac{1}{2}\chi^{\alpha \beta}\partial_{\beta}\Gamma)=\braket{\nabla \Gamma,D+\frac{1}{2}\chi\nabla \Gamma}=0.
\end{align}
It is solved implicitly by decomposing the deterministic force $\b{D}$ into the two orthogonal fields $\nabla \Gamma$ and $\b{r}$ as
\begin{align}\label{EQ:DriftDecomposition}
D^{\alpha}=-\frac{1}{2}\chi^{\alpha \beta}\partial_{\beta}\Gamma+r^{\alpha}. 
\end{align}
The orthogonality condition $r^{\alpha} \partial_{\alpha}\Gamma=0$ demands that $\b{r}$ encodes dynamics on equipotential surfaces of $\Gamma$. The first term $(-1/2\chi^{\alpha \beta}\partial_{\beta}\Gamma)$ is responsible for the stability of the fixed points of the deterministic force $\b{D}$. The transversal decomposition implies that $\nabla \Gamma \neq -2 \b{\chi}^{-1}\b{D}$, such that in general it is an impossible task to obtain non-perturbative analytical expressions for the non-equilibrium potential $\Gamma$ in systems lacking detailed balance \cite{Bouchet2016}. The saddle-point trajectories \eqref{EQ:DeterministicSaddlePoint2} describing noise-activation  are
\begin{align}\label{EQ:Instanton}
\partial_{t}\sigma^{\alpha}&=D^{\alpha}+\chi^{\alpha \beta} \partial_{\beta}\Gamma=\frac{1}{2}\chi^{\alpha \beta}\partial_{\beta}\Gamma+r^{\alpha}=-D^{\alpha}+2r^{\alpha},\\
\tilde{\sigma}^{\alpha}&=\partial_{\alpha}\Gamma. 
\end{align}
A comparison with the deterministic dynamics  $\partial_t \sigma^{\alpha}=D^{\alpha}$ shows that in general the noise activated trajectories are not the time-reversed partners of the deterministic dynamics and visit different regions in the Bloch sphere as seen in Fig.\,(2). Both deterministic and noise-activated trajectories share, however, the same set of fixed points $\{\b{\sigma_0}\}$ with $D(\{\b{\sigma_0}\})=0$. Since all fixed points are hyperbolic, there exists one stable ($M_s$) and one unstable ($M_u$) 3-dimensional manifold for the zero-energy Hamilton that intersect at the set of fixed points $\{\b{\sigma_0}\}$. The stable manifold $M_s$ is characterized by $\b{\tilde{\sigma}}=0$ and $\partial_t \b{\sigma}=\b{D}$ and the unstable manifold $M_u$ is characterized by $\tilde{\b{\sigma}}(\b{\sigma})=\nabla \Gamma (\b{\sigma})$ and $\partial_t \b{\sigma}=-\b{D}+2\b{r}$.
The cost for a fluctuation is then measured by the action 
\begin{align}
S^{\rm fluc}(\sigma)=N\int_{t_0}^t \b{\tilde{\sigma}} \partial_t \b{\sigma} =N\int_{\sigma_0}^{\sigma}\b{\tilde{\sigma}}d\b{\sigma}.
\end{align}
In the weak noise limit $(N \to \infty)$ one can define a non-equilibrium potential or cost function as the minimal action acquired for a path connecting a fixed point $\b{\sigma}_0$ to any given point $\b{\sigma}$ on the manifold $M_u$ as
\begin{align} 
\psi(\b{\sigma}_0,\b{\sigma})={\rm min}\{S^{\rm fluc}_{[t_0,t],\sigma_0}(\sigma):\sigma(t_0)=\sigma_0,\sigma(t)=\sigma,t_0<t\}.
\end{align}
The result of the integration is independent of the path $(\b{\sigma}(t),\b{\tilde{\sigma}}(t))$ in $M_u$ when $M_u$ is locally defined by $\b{\tilde{\sigma}}=\nabla \Gamma$. To exponential accuracy the probability $p$ for a noise-induced trajectory is then given by $p\sim \exp\left(-N \psi(\b{\sigma}_0,\b{\sigma})\right)$. This explains the \textit{first} point that the mean occupation times $\braket{t_{Sr,0}}\sim A_{Sr,0}\exp(N\psi_{Sr,0})$. For weak, but finite noise, the instanton trajectory is given by $\partial_t \b{\sigma}=-\b{D}+2\b{r}+\frac{1}{\sqrt{N}} \b{\xi}$ and spreads around the deterministic instanton path as, observed in Fig.\,(2).

We turn our attention to the \textit{second} point and check violation of detailed balance. Under microscopic reversibility \cite{Graham80,Graham71} the deterministic force $D$ and diffusive contribution $\chi$ in general obey a relation fixed by \cite{Graham70}
\begin{align}\label{EQ:DetailedBalance}
\text{rot}[\b{\chi}^{-1}(2\b{D}-\nabla\b{\chi})]=0,
\end{align}
which is not satisfied for the stochastic optical Bloch equations (15). We conclude the absence of detailed balance for the dynamics studied that we present in this work.

Coming to the third point, for out-of-equilibrium systems in general $\b{D}$ itself is not a gradient field and has both a non-conservative $(\nabla \times D \neq 0)$ and a conservative contribution $(\nabla \cdot D \neq 0)$. Whenever the generators for conservative and non-conservative dynamics do not commute, i.e. when the corresponding trajectories are not orthogonal, both fields $(-\frac{1}{2}\chi^{\alpha \beta}\partial_{\beta}\Gamma)$ and $r^{\alpha}$ contribute to both conservative and non-conservative parts of $\b{D}$. Thus, having non-conservative and conservative dynamics is necessary to reach a non-equilibrium steady state but not sufficient, the counter example being a spin subject to a magnetic field $\b{B}=\mu B \b{e}_z$ and spontaneous emission. 

The presence of non-conservative and conservative forces together with the absence of detailed balance allows the conclusion that the steady-state for the stochastic optical Bloch equations in this work is firmly out-of equilibrium. \\

\section{Outlook}
Extending the Dicke model to the dissipative regime by adding cooperative losses drastically enriches the dynamics, introducing a bicritical point as well as a bistable, fluctuation dominated regime for the superradiance transition. This defines a framework to study a whole set of dissipative phase transitions both in theory and experiment and should motivate current experiments to explore the dissipative Dicke realm \cite{Zhiqiang17, MBarrett17}. 
While a brief calculation with current experimental parameters shows that the cooperative loss rate typically is too weak to set it in competition to the effective atom-atom coupling strength $J$, it reveals that short effective distances, i.e. large densities of atoms will be able to access this regime, which might encourage experimental progress in this direction. 

Possible experimental observations include fluctuation induced lasing and dynamical hysteresis when the atom-light coupling is swept between the two spinodal lines \cite{rodriguez2017}. Slightly lifting the symmetry protected degeneracy between the two superradiant states introduces three, genuinely different metastable states. This configuration enables the study of dynamics in the absence of detailed balance on macroscopic scales that are exclusive to nonequilibrium systems, e.g. experimentally accessible via circulating currents in the stationary state.

\acknowledgements The authors thank S. Diehl and A. Rosch for inspiring discussions and comments on the manuscript. J.G. thanks the Bonn-Cologne Graduate School of Physics and Astronomy (BCGS) for financial support. M.B. acknowledges support from the Alexander von Humboldt foundation and funding by the German Research Foundation (DFG) through the Institutional Strategy of the University of Cologne within the German Excellence Initiative (ZUK 81).


\appendix
\section{Derivation of Noise Operators and Noise Correlations in a Heisenberg-Langevin-Framework}
\label{App:MakrovBath}
The appearance of the noise-operator $\xi^{O_i}$ in the Heisenberg-Langevin equation for the operator $O_i$ results from the fluctuation-dissipation relations of the eliminated bath. It ensures the correct time evolution of fluctuations $O_iO_\ell$ and preserves operator commutation relations for open systems in time \cite{Scully97}. Their explicit form is derived from the unitary Heisenberg equation of motion for $O_i$. The noise kernel $\chi_{i\ell}(\sigma)=\braket{\xi^{O_i}\xi^{O_\ell}}_{\rm bath}$ is then computed by evaluating the bath operators in the Born-Markov approximation. We apply the standard Heisenberg-Langevin theory where the interaction of the system with the external bath is specified in terms of a Hamiltonian $H_{\rm sys-bath}$ that couples the bath modes linearly to the system operators, see e.g. \cite{Scully97}. The two statistically independent baths for the photons and for the atoms are the continuum of radiation modes outside the cavity. 
We consider the system-bath Hamiltonians in the interaction picture 
\begin{align}
H^{\rm atoms}_{\rm bath-sys}(t)&=\sum_{k,\ell=1}^N \left(\varepsilon_{k,\ell}\sigma^+_{\ell}b_k e^{i(\omega_z-\nu_k)t}+cc.\right),\\
H^{\rm photons}_{\rm bath-sys}(t)&=\sum_{k} \left(\tilde{\varepsilon}_k a^{\dagger}c_k e^{i(\omega_0-\omega_k)t}+cc.\right),
\end{align}
where we have coupled all emitters $\sigma^{\pm}_{\ell}$ in $H^{\rm atoms}_{\rm bath-sys}(t)$ to one set of bath modes $b^{\dagger}_k b_k$ which also allows for collective emission of excitations outside of the cavity that become relevant in the superradiant regime. Here $\omega_0$ and $\omega_z$ refer to effective frequencies for the photons and the atoms in a frame rotating at a frequency set by the external laser drive, see e.g. \cite{Dimer07}.
The Heisenberg equation of motions for the system and the bath operators can be written as
\begin{align}
\partial_t a_t&=-i[a_t,H_t]=-i\sum_k \tilde{\varepsilon}_k c_{k,t}e^{i(\omega_0-\omega_k)t}\label{EQ:SystemBath0},\\
\partial_t b_{k,t}&=-i[b_{k,t},H_t]=-i\sum_{\ell=1}^N \varepsilon^{*}_{k,\ell}\sigma^{-}_{\ell,t}e^{-i(\omega_z-\nu_k)t},\\
\partial_t c_{k,t}&=-i[b_{k,t},H_t]=-i\sum_{\ell=1}^N \tilde{\varepsilon}^{*}_{k,\ell}a_t e^{-i(\omega_z-\omega_k)t},\\
\partial \sigma^{-}_{\ell',t}&=-i[\sigma^{-}_{\ell',t},H_t]=i\sum_{k}\sigma^{z}_{\ell',t}b_{k,t}e^{i(\omega_z-\nu_k)t}\varepsilon_{k,\ell}\label{EQ:SystemBath1},\\
\partial \sigma^{z}_{\ell',t}&=-i[\sigma^{z}_{\ell,t},H_t]=\sum_k\left(-2i\sigma^{+}_{\ell',t}b_{k,t}e^{i(\omega_z-\nu_k)t}\varepsilon_{k,\ell'} -cc.\right)\label{EQ:SystemBath2}.
\end{align}
Here we have used $H_t=H^{\rm atoms}_{\rm bath-sys}(t)+H^{\rm photons}_{\rm bath-sys}(t)$. We eliminate the bath degree of freedom by formal integration of their equations of motion
\begin{align}
b_{k,t}=b_{k,0}-i\int_{0}^t dt' \sum_{\ell=1}^N \varepsilon^{*}_{k,\ell}\sigma^{-}_{\ell,t'}e^{-i(\omega_z-\nu_k)t'}\label{EQ:BathModeAtom},\\
c_{k,t}=c_{k,0}-i\int_{0}^t dt' \sum_{\ell=1}^N \tilde{\varepsilon}^{*}_{k,\ell}a_{t'}e^{-i(\omega_z-\omega_k)t'}
\end{align}
and insert Eqs.~(\ref{EQ:BathModeAtom}) and the conjugates into the equations of motion for the system operators given by Eqs.~(\ref{EQ:SystemBath0}-\ref{EQ:SystemBath2}). 
\begin{align}
\partial_t a_t&=\xi^{a}_t-\int_{0}^t dt' \sum_k |\tilde{\varepsilon}_k|^2 \mathcal{F}^{*}_{k,\omega_0,t,t'}a_{t'}\label{EQ:DissA},\\
\partial_t \sigma^{+}_{\ell',t}&=\xi^{+}_{\ell',t}+\int_{0}^t dt' \sum_{\ell,k} |\varepsilon_k|^2 e^{i(\b{k}-\b{k_0})(\b{r_{\ell}}-\b{r_{\ell'}})}\mathcal{F}_{k,\omega_z,t,t'}\sigma^{+}_{\ell,t'}\sigma^z_{\ell',t},\\
\partial_t \sigma^{-}_{\ell',t}&=\xi^{-}_{\ell',t}+\int_{0}^t dt' \sum_{\ell,k} |\varepsilon_k|^2 e^{-i(\b{k}-\b{k_0})(\b{r_{\ell}}-\b{r_{\ell'}})}\mathcal{F}^{*}_{k,\omega_z,t,t'}\sigma^z_{\ell',t}\sigma^{-}_{\ell,t'},\\
\partial_t \sigma^z_{\ell',t}&=\xi^z_{\ell',t}-2\left(\int_{0}^t dt' \sigma^{+}_{\ell',t}\sum_{\ell}|\varepsilon_k|^2 e^{i(\b{k}-\b{k_0})(\b{r_{\ell}}-\b{r_{\ell'}})} \mathcal{F}_{k,\omega_z,t,t'} \sigma^{-}_{\ell,t'}+cc.\right)\label{EQ:DissZ}.
\end{align}
Equations \eqref{EQ:DissA}-\eqref{EQ:DissZ} now contain only the dissipative and fluctuating components that arise from the interaction of the system with the external reservoir. The explicit form of the noise-operators can be read off as
\begin{align}
\xi^{a}_t&=-i\sum_k \tilde{\varepsilon}_k c_k(0)e^{-i(\omega_0-\omega_k)t}\label{EQ:PhotonNoiseOperator},\\
\xi^{+}_{\ell',t}&=-i \sum_{k}b^{+}_{k}(0)\sigma^z_{\ell'}(t)e^{-i(\omega_z-\nu_k)t}\varepsilon^{*}_k e^{i(\b{k}-\b{k_0})\b{r}_{\ell'}}\label{EQ:NoiseAtomsx},\\
\xi^{-}_{\ell',t}&=i \sum_{k}\sigma^z_{\ell',t}b_{k}(0)e^{i(\omega_z-\nu_k)t}\varepsilon_k e^{-i(\b{k}-\b{k_0})\b{r}_{\ell'}},\\
\xi^{z}_{\ell ',t}&=\sum_{k}\left(2i b^{+}_{k}(0)\sigma^{-}_{\ell',t}\varepsilon^{*}_k e^{-i(\omega_z-\nu_k)t}e^{i(\b{k}-\b{k_0})\b{r}_{\ell'}}+ cc.\right)\label{EQ:NoiseAtomsz},
\end{align}
within the Born-Markov approximation, the frequency independent damping constants are parametrised by the relations 
\begin{align}
\gamma\delta(t-t')&=\sum_k |\varepsilon_k|^2 \mathcal{F}_{k,t,t'}=2\pi |\varepsilon_{\omega_z}|^2\mathcal{D}(\omega_z)\delta(t-t'),\\
\kappa\delta(t-t')&=\sum_k |\tilde{\varepsilon}_k|^2 \mathcal{F}_{k,t,t'}=2\pi |\varepsilon_{\omega_0}|^2\mathcal{D}(\omega_0)\delta(t-t'),\nonumber\\
\mathcal{F}_{k,\omega,t,t'}&=\exp[-i(\omega-\nu_k)(t-t')].
\end{align}
Here, we have taken $\varepsilon_{k,\ell}=\varepsilon_k e^{-i(\b{k}-\b{k}_0)\b{r}_{\ell}}$ as the cavity-shifted, spatially dependent atom-photon coupling to the bath modes $b_k$ outside of the cavity, where $\b{k_0}$ is the cavity wave vector. $\mathcal{D}(\omega_0)$ and $\mathcal{D}(\omega_z)$ are the density of states of the bath modes and $\varepsilon_{\omega_0},\varepsilon_{\omega_z}$ are the microscopic system-bath coupling constants evaluated at the effective photon frequency $\omega_0$ and the effective atom frequency $\omega_z$. 


\section{Elimination of Cavity-Photons in the presence of Noise}
{\label{App:AdiabaticElimination}}
We detail the elimination of the  cavity photons in the presence of photonic noise functions $(\xi^a,\xi^{a^{\dagger}})$. As a result, the local atomic components ($\sigma^x_i,\sigma^y_i,\sigma^z_i$) at site $i$ inherit additional noise from the photons with strength $\propto \frac{2\kappa J}{\omega_0}$, where $J=4\frac{g^2 \omega_0}{\omega^2_0+\kappa^2}$. 
Since the photons mediate an all-to-all coupling of the atoms for the deterministic dynamics, the noise inherited from the photons adds to the collective loss term of the atoms $\sim \gamma \beta$ in the $\sigma^y$ and $\sigma^z$-channel. However, since both co- and counter-rotating terms $\sim (a+a^{\dagger})\sum_{\ell}\sigma^x_{\ell}$ are present in the Dicke-Hamiltonian, the $\sigma^{x}$-channel does not inherit a photonic noise component.
We start by considering the Heisenberg-Langevin equations given in Eq.\,(6)-(9)
\begin{align}
\partial_t a&=-\l\kappa+i\omega_0\r a-i \frac{g}{\sqrt{N}}\sum_{\ell=1}^N \sigma^x_\ell+\xi^a\label{App:MeanFieldSR1}\;,\\
\partial_t \sigma^y_i&=\omega_z\sigma^x_i -\left[\frac{2g(a^\dagger+a)}{\sqrt{N}}-\frac{\gamma \beta}{N}\hspace{-0.1cm}\sum_{\ell\neq i} \sigma^{y}_{\ell}\right]\sigma^z_i-\gamma\sigma^{y}_i+\xi^y_i\;,\\
\partial_t \sigma^z_i&=\frac{2g(a^{\dagger}+a)}{\sqrt{N}}\sigma^y_i-2\gamma(1+\sigma^z_i)+\xi^z_i \nonumber\\
&-\frac{\gamma\beta}{2N} \sum_{\ell\neq i}\l \sigma^{x}_{i}\sigma^{x}_{\ell}+\sigma^{y}_{i}\sigma^{y}_{\ell}+i(\sigma^y_{i}\sigma^{x}_{\ell}-\sigma^{x}_{i}\sigma^{y}_{\ell})+cc. \r. 
\end{align}
On the level of single operator expectation values for the system variables $\braket{.}\equiv\braket{.}_{\rm sys}$ one can define a collective variable as $\sigma^{\beta}=\sum_{\ell=1}^N\braket{\sigma^\beta_{\ell}}/N$, the collective atom noise $\xi^\beta=\sum_\ell\langle\xi^\beta_\ell\rangle/N$, the noise function of the photons as $\eta= \braket{\xi^a}/\sqrt{N}$ and the expectation value of the photon operator $\alpha=\braket{a}/\sqrt{N}$, (noise operators vanish only for bath averaging)
\begin{align}
\partial_t \alpha&=-\l\kappa+i\omega_0\r \alpha-ig \sigma^x+\eta\label{App:MeanFieldSR1}\;,\\
\partial_t \sigma^y&=\omega_z\sigma^x-\left[2g(\alpha+\alpha^{*})-\gamma \beta \sigma^{y} \right]\sigma^z-\gamma\sigma^{y}+\xi^y\label{EQ:SigmaYComponent}\;,\\
\partial_t \sigma^z&=2g(\alpha+\alpha^{*})\sigma^y-2\gamma(1+\sigma^z)
-\frac{\gamma\beta}{2} [\sigma^x\sigma^x+ \sigma^y\sigma^y]+\xi^z. \label{EQ:SigmaZComponent}
\end{align}
We eliminate the gapped photon degrees of freedom $(\partial_t \alpha=0)$ to obtain their steady-state value as
\begin{align}
\alpha+\alpha^{*}=\frac{-\eta+i g \sigma^x}{-\kappa -i\omega_0}+\frac{-\eta^*-ig \sigma^x}{-\kappa +i\omega_0}.\label{EQ:SteadyStatePhoton}
\end{align}
Plugging Eq.\,\eqref{EQ:SteadyStatePhoton} in Eq.\,\eqref{EQ:SigmaYComponent} and in Eq.\,\eqref{EQ:SigmaZComponent} leads to a redefined noise function in the $\sigma^{y(z)}$-channel
\begin{align}
\tilde{\xi}^{y(z)}&=\xi^{y(z)}\mp\frac{2g}{\kappa^2+\omega^2_0}\sigma^{z(y)}\bigg[\kappa(\eta+\eta^*)-i\omega_0(\eta-\eta^*)\bigg] \label{EQ:ModifiedAtomicNoise}\\
\partial_t \sigma^y&=\omega_z\sigma^x+J\sigma^x \sigma^z+\gamma \beta \sigma^{y} \sigma^z-\gamma\sigma^{y}+\tilde{\xi}^y\\
\partial_t \sigma^z&=-J\sigma^x\sigma^y-2\gamma(1+\sigma^z)
-\frac{\gamma\beta}{2} [\sigma^x\sigma^x+ \sigma^y\sigma^y]+\tilde{\xi}^z
\end{align}

The covariances of the atoms now contain a noise contribution from the photon field
\begin{align}
\braket{\tilde{\xi}^{y(z)}\tilde{\xi}^{y(z)}}_{\rm noise}&=\braket{\xi^{y(z)}\xi^{y(z)}}_{\rm noise}+ \sigma^{z(y)}\sigma^{z(y)} \frac{J}{\omega_0}\braket{\eta^*\eta+\eta\eta^*}_{\rm noise},\label{EQ:RedefinedNoisey}\\
\braket{\tilde{\xi}^{y}\tilde{\xi}^{z}}_{\rm noise}&=\braket{\xi^{y}\xi^{z}}_{\rm noise}-\sigma^{z}\sigma^{y} \frac{J}{\omega_0}\braket{\eta^*\eta+\eta\eta^*}_{\rm noise}.\label{EQ:RedefinedNoisez}
\end{align}
Taking the bath for the photons to be in a zero temperature vacuum state, the noise correlation function is according to Eq.\,\eqref{EQ:PhotonNoiseOperator}
\begin{align}
\langle \eta^*\eta+\eta\eta^*\rangle=2\kappa/N\delta(t-t').
\end{align}
We remark that the so-obtained variances for the atoms and photons are equivalent to the variances that would be obtained in the associated MSRDJ-path integral \cite{MSR1973,Janssen1976,Dominicis1978} for the complex fields $(a_t,a^*_t)$ and for the real fields $(\sigma^x_t,\sigma^y_t,\sigma^z_t)$ where the photon degrees of freedom are then integrated out exactly for the zero-frequency sector.
\section{Derivation of the classical Noise Kernel $\b{\chi}$ for stochastic optical Bloch equations}
\label{App:ClassicalNoiseKernel}
The classical noise-kernel $\b{\chi}(\sigma)$, see Eq.\,(17) contains the noise correlations for the collective atomic variables $(\sigma^{\alpha}=\sum_{\ell=1}^N \sigma^{\alpha}_{\ell}/N)$ in the stochastic optical Bloch equations. It is derived by mapping the corresponding operator-valued noise correlations of the atomic and photonic noise functions to correlations of the associated classical noise functions through the process of symmetrisation. We detail the derivation of the classical noise kernel and start by explicitly evaluating the operator-valued atomic and photonic noise functions in Eqs.~(\ref{EQ:NoiseAtomsx}-\ref{EQ:NoiseAtomsz}). 
\subsection{Evaluation of operator-valued noise correlations in the atomic channel}
The evaluation of noise correlation functions is now performed as an average over the bath degrees of freedom denoted as $\braket{.}_{\rm bath}$ where the expectation value is a thermal average over a zero-temperature bath. Since the external bath is described by the vacuum and only terms $\propto b_k(0)b_k^{\dagger}(0)$ contribute, correlations of the form $\braket{\xi^{+}(t)\dots}=0$ and $\braket{\dots \xi^{-}(t)}=0$ vanish. In particular that means $\braket{\xi^{+}(t)\xi^{-}(t)}=0$ whereas $\braket{\xi^{-}(t)\xi^{+}(t)}\neq 0.$ 

The noise-correlations for the atomic degrees of freedom can be expressed by using Eqs.~(\ref{EQ:NoiseAtomsx}-\ref{EQ:NoiseAtomsz}) as
\begin{align}
\braket{\xi^{i}_{\ell',t'}\xi^{j}_{\ell,t}}_{\rm bath}&=\gamma\delta(t-t')\bigg[\delta_{\ell,\ell'}+(1-\delta_{\ell,\ell'})\alpha\bigg]\tilde{\chi}^{ij}_{\ell'\ell}\label{EQ:NoiseRelation},\\
\tilde{\chi}^{ij}_{\ell' \ell}&=\left(
\begin{array}{ccc}
 0 & 0 & 0 \\
 \sigma^{z}_{\ell',t}\sigma^z_{\ell,t} & 0 & -2\sigma^{z}_{\ell',t}\sigma^{-}_{\ell,t} \\
  -2\sigma^{+}_{\ell',t}\sigma^{z}_{\ell,t} & 0 & 4 \sigma^{+}_{\ell',t}\sigma^{-}_{\ell,t} \\
\end{array}
\right)_{ij}
\label{EQ:NoiseOperatorsCorrelsAtom}
\end{align}
where the indices $(i,j)\in(+,-,z)$ refer to the atomic variables. The entries of the matrix $\tilde{\chi}^{ij}_{\ell' \ell}$ are still dependent on the system operators $\sigma^{i}_{\ell}$ at site $\ell$. 
With the relations $\xi^x_{\ell}=\xi^{+}_{\ell}+\xi^{-}_{\ell}$ and $\xi^y_{\ell}=-i (\xi^{+}_{\ell}-\xi^{-}_{\ell})$ we rotate from $(+,-,z)$ into the $(x,y,z)$ basis
\begin{align}
\tilde{\chi}^{ij}_{\ell' \ell}=
\left(
\begin{array}{ccc}
 \sigma^{z}_{\ell'}\sigma^{z}_{\ell} & - i \sigma^{z}_{\ell'}\sigma^{z}_{\ell} & -\sigma^{z}_{\ell'}(\sigma^{x}_{\ell}-i\sigma^{y}_{\ell}) \\
 i \sigma^{z}_{\ell'}\sigma^{z}_{\ell} & \sigma^{z}_{\ell'}\sigma^{z}_{\ell} & -i \sigma^{z}_{\ell'}(\sigma^{x}_{\ell'}-i\sigma^{y}_{\ell'}) \\
 -(\sigma^{x}_{\ell'}+i\sigma^{y}_{\ell'})\sigma^{z}_{\ell} & i (\sigma^{x}_{\ell'}+i\sigma^{y}_{\ell'})\sigma^{z}_{\ell} & \tilde{\chi}^{zz}_{\ell \ell} \\
\end{array}
\right)_{ij},
\label{EQ:MatrixMllprime}
\end{align}
with $\tilde{\chi}^{zz}_{\ell\ell'}=\l \sigma^{x}_{\ell'}\sigma^{x}_{\ell}+\sigma^{y}_{\ell'}\sigma^{y}_{\ell}+i(\sigma^y_{\ell'}\sigma^{x}_{\ell}-\sigma^{x}_{\ell'}\sigma^{y}_{\ell})\right)$.
For the components with $\ell=\ell'$ the local spin algebra can be used to write the correlation matrix for the local noise components as
\begin{align}
\tilde{\chi}^{ij}_{\ell \ell}&=\left(
\begin{array}{ccc}
 1 & -i & (\sigma^{x}_{\ell}-i\sigma^{y}_{\ell}) \\
 i & 1 & i (\sigma^{x}_{\ell}-i\sigma^{y}_{\ell}) \\
 (\sigma^{x}_{\ell}+i\sigma^{y}_{\ell}) & -i (\sigma^{x}+i\sigma^{y}_{\ell}) & 2(1+\sigma^{z}_{\ell}) \\
\end{array}
\right)_{ij}\label{EQmxyzbasis}.
\end{align}
Equations \eqref{EQ:MatrixMllprime}-\eqref{EQmxyzbasis} together with Eq.\,\eqref{EQ:ModifiedAtomicNoise} are the starting point to obtain the corresponding classical noise kernel $\chi(\sigma)$. 
\subsection{Mapping the photonic and atomic noise correlations to a classical noise kernel}
The general correlation matrix $\chi_{\ell \ell'}$ in Eq.\,\eqref{EQ:MatrixMllprime} was obtained by averaging over the bath degrees of freedom and still depends on the system operators $\sigma^{\alpha}_\ell$ with $\alpha=(x,y,z)$ and $\ell$ as the local site index. For a mapping to a classical noise correlation matrix it is necessary to erase the information on commutation relations. This is achieved by symmetrising the matrix entries $\chi^{ij}_{\ell \ell'}$ which amounts to taking their real part $\text{Re}[\chi^{ij}_{\ell \ell'}]$. As described in the main text, this procedure leads to a symmetric, real and positive definite noise kernel and is thus well-defined. \\
We are interested in the noise strength for the equations of motion of the collective variable $\sigma^{\alpha}=\sum_{\ell}\braket{\sigma^{\alpha}_{\ell}}_{\rm sys}/N$ with the averaged noise $\tilde{\xi}^{\alpha}=\sum_{\ell} \braket{\tilde{\xi}^{\alpha}_{\ell}}_{\rm sys}/\sqrt{N}$, where $\braket{.}_{\rm sys}$ is a quantum mechanical average over the system variables. Here $\tilde{\xi^{\alpha}}$ is the modified noise function of the atoms that contains both an atomic and photonic contribution for $\alpha=(y,z)$ stemming from the elimination of the cavity degrees of freedom and is defined in Eq.\,\eqref{EQ:ModifiedAtomicNoise}. The $\sigma^x$-channel is free of a photonic contribution as discussed previously. The noise average $\braket{.}_{\rm noise}$ to obtain the classical correlation matrix $\b{\chi}$ is then defined by averaging over both bath and system degrees of freedom
\begin{widetext}
\begin{align}
\langle \tilde{\xi}^{\alpha}\tilde{\xi}^{\beta}\rangle_{\text{noise}}=
N^{-1}\langle\sum_{\ell,m}\tilde{\xi}^{\alpha}_{\ell}\tilde{\xi}^{\beta}_{m}\rangle_{\text{sys+bath}}=N^{-1}\delta(t-t')\sum_{\ell m}\left(\gamma\bigg[\delta_{\ell,m}+(1-\delta_{\ell,m})\alpha\bigg]\text{Re}[\braket{\tilde{\chi}^{\alpha \beta}_{\ell m}}_{\rm sys}]+N^{-1}\frac{2J\kappa}{\omega_0}\braket{M^{\alpha \beta}_{\ell m}}_{\rm sys}\right)
=\delta(t-t')\chi^{\alpha \beta}(\b{\sigma}),
\label{EQ:DerivationNoiseCorrelationMatrix}
\end{align}
\end{widetext}
where the contributions from the photons is specified as
\begin{align}
\braket{M^{yy}_{ij}}_{\rm sys}&=\braket{\sigma^{z}_i}\braket{\sigma^{z}_j}, \ \ \ \braket{M^{zz}_{ij}}_{\rm sys}=\braket{\sigma^{y}_i}\braket{\sigma^{y}_j},\\
\quad \braket{M^{yz}_{ij}}_{\rm sys}&=\braket{M^{yz}_{ij}}_{\rm sys}=-\braket{\sigma^{z}_i}\braket{\sigma^{y}_j}.
\end{align}
The noise correlation matrix $\chi^{\alpha \beta}(\b{\sigma})$ is specified in Eq.\,(17) and contains now a noise component from the local, uncorrelated loss processes $\sim \gamma$ as well as from the collective loss processes $\sim \gamma \beta=\gamma \alpha (N-1)$, where $\beta=const.$ in the thermodynamic limit and a contribution from the photons $\sim 2J\kappa/\omega_0$. From Eq.\,\eqref{EQ:DerivationNoiseCorrelationMatrix} one can see that the variance of the sum of the random noise functions $\sum_{\ell} \braket{\tilde{\xi}^{\alpha}_{\ell}}_{\rm sys}$ scales with the number of atoms $N$ as expected for instance from the central limit theorem. 
\onecolumngrid
\section{Non-local Lindblad contribution in Born-Markov approximation}
\label{App:SecCollectiveAtomicDecay}
For the derivation of the collective decay contribution, we will briefly review the textbook approach (see e.g.\,\cite{Steck}) of how the external reservoir influences the evolution of the system in a Born-Markov approximation. This leads to a in general non-local density matrix equation, see \eqref{EQ:GeneralLindbladNonLocalDensityMatrix}. Collective decay contributions that add to the single atom decay rates have been derived in the context of single photon sub- and superradiant states \cite{Scully2015} in a wave-function formalism. Here, we carry these considerations over to a density matrix formalism and show that a description in terms of Lindblad operators reproduces the results obtained from the wave function picture. The collective loss contribution emerges by allowing all spins to interact with one shared bath. 
In the interaction picture, the system-bath Hamiltonian in the rotating wave approximation can be written as

\begin{align}
H^{\rm atoms}_{\rm bath-sys}(t)&=\sum_{k}\sum_{\ell=1}^N \left(\varepsilon_{k,\ell}\sigma^+_{\ell}b_k e^{i(\omega_z-\nu_k)t}+\varepsilon^{*}_{k,\ell} b^{\dagger}_k\sigma^{-}_{\ell} e^{-i(\omega_z-\nu_k)t}\right).
\end{align}

Here, the coupling to the bath is given by $\varepsilon_{k,\ell}=\varepsilon_k e^{-i \b{k} \b{r}_{\ell}}$, with $\varepsilon_k$ taking into account the frequency dependence of the $k$th radiation mode that is given by $\nu_k=ck$. The time-evolution of the full system and bath density matrix reads $(\hbar=1)$
\begin{align}
\partial_t \rho(t)_{\rm}&=-i\bigg[H^{\rm atoms}_{\rm bath-sys}(t),\rho(0)-i\int_{0}^t dt' \bigg[H^{\rm atoms}_{\rm bath-sys}(t'),\rho(t')\bigg]\bigg].
\label{EQ:DensityMatrix}
\end{align}
For weak system-reservoir coupling in the Born-Markov approximation, the density matrix is written as
\begin{align} 
\rho(t')\equiv \rho_{\rm bath-sys}(t')\approx \rho_{\rm sys}(t')\otimes \rho_{\rm bath}(0)+\delta \rho_{\rm bath-sys}(t'),
\end{align}
where the last term is of order $\mathcal{O}(\varepsilon_k)$. This is justified for a large reservoir, which is unaffected by the system dynamics and for which the bath-system coupling is memoryless, i.e. $\rho(t')\rightarrow \rho(t)$ in Eq.~\eqref{EQ:DensityMatrix}. 

Tracing out the bath degrees of freedom in the Born-Markov approximation, the time evolution for the system is found to be \begin{align}
\partial_t \rho_{\rm sys}(t)=&\int_0^t dt' \sum_{k,\ell,\ell'}|\varepsilon_{k}|^2 e^{-i(\b{k}-\b{k}_0)(\b{r}_{\ell'}-\b{r}_{\ell})} \bigg[\sigma^{-}_{\ell'}\rho_t\sigma^{+}_{\ell}\left(\zeta_{k}+\zeta^{*}_{k}\right)-
\rho_t\sigma^{+}_{\ell'}\sigma^{-}_{\ell}\zeta^{*}_{k}-
\sigma^{+}_{\ell'}\sigma^{-}_{\ell}\rho_t\zeta_{k}\bigg],\label{EQ:GeneralLindbladNonLocalDensityMatrix}
\end{align}
where we have made explicit that in the Dicke model all momenta are expressed with respect to the cavity wave vector $\b{k}_0$ and we have collected temporal phase factors as $\zeta_{k}(t'-t)=\exp\left(-i(\omega_z-\nu_k)(t'-t)\right)$. 

For a large number of atoms, the sum over all atoms and momenta is only non-vanishing for two different contributions. Either for $\ell=\ell'$, which describes the uncorrelated, single atom decay. Or for  $|\b{k}-\b{k}_0|\approx 0$, which describes correlated decay into modes near the cavity wave vector. We note that contributions with $\ell\neq\ell'$ are generally suppressed by the volume factor $\propto 1/V$, which is implicit in the atom-light coupling constant $\varepsilon_k$. We single out the uncorrelated single atom loss, which has been treated in many previous works (e.g. see \cite{Scully97}) and find
\begin{align}
\partial_t \rho_{\rm sys}(t)=&\gamma\sum_{\ell=1}^N \left(\sigma^{-}_{\ell}\rho_t\sigma^{+}_{\ell}-\frac{1}{2}\{\sigma^{+}_{\ell}\sigma^{-}_{\ell},\rho_t\}\right)\nonumber\\
&+\int_0^t dt' \sum_{k,\ell'\neq\ell}e^{-i(\b{k}-\b{k}_0)(\b{r}_{\ell'}-\b{r}_{\ell})}|\varepsilon_{k}|^2\left( \sigma^{-}_{\ell'}\rho_t\sigma^{+}_{\ell}\left(\zeta_{k}+\zeta^{*}_{k}\right)-\rho_t\sigma^{+}_{\ell'}\sigma^{-}_{\ell}\zeta^{*}_{k}-\sigma^{+}_{\ell'}\sigma^{-}_{\ell}\rho_t\zeta_{k}\right)\label{EQ:NonLocalLindbladContribution}
\end{align}   

For the second contribution to Eq.~\eqref{EQ:NonLocalLindbladContribution}, we focus only on the collective part that arises from wave vectors $|\b{k}-\b{k}_0|\approx 0$. We proceed by calculating the weight of the associated delta function
\begin{align}
&\int_0^t dt' \sum_{k} |\varepsilon_k|^2 e^{-i(\b{k}-\b{k}_0)(\b{r}_{\ell'}-\b{r}_{\ell})}\zeta_k \approx \int_0^t dt' \sum_{k}|\varepsilon_k|^2 \frac{(2\pi)^3}{V}\delta(\b{k}-\b{k}_0)\zeta_k
\end{align}
We assume that the bath modes lie dense and work in the continuum limit to use the replacements
\begin{align}
\delta(\b{k}-\b{k}_0)&=\frac{1}{2\pi}\int_{-R}^R e^{i(k-k_0)r}\delta(\theta_k-\theta_{k_0})\delta(\phi_k-\phi_{k_0})dr\frac{1}{k^2 \sin(\theta_k)}\\
\sum_k &\to \frac{V}{(2\pi)^3}\int_0^{\infty} dk k^2 \int_{0}^{\pi}\sin(\theta_k) d\theta_k \int_0^{2\pi} d\phi_k.
\end{align}
which leads to the integral
\begin{align}
\int_0^t dt' \sum_{k} |\varepsilon_k|^2 e^{-i(\b{k}-\b{k}_0)(\b{r}_{\ell'}-\b{r}_{\ell})}\zeta_k \approx& \int_0^t dt' \int_0^\infty dk |\varepsilon_k|^2 \bigg[\frac{1}{2\pi}\int_{-R}^R e^{-i(k-k_0)r}dr\bigg] e^{i(ck-\omega_z)(t-t')}\nonumber\\
&= \int_0^t dt' \int_0^\infty dk |\varepsilon_k|^2 \frac{1}{2\pi}\int_{-R}^R \exp\bigg[ic(k-k_0)(t-t'-r/c)+i c(k_0-k_z)(t-t')\bigg] dr \nonumber \\
&= |\varepsilon_{k_0}|^2\int_0^t dt'\int_{-R}^R dr \frac{1}{2\pi}\frac{2\pi}{c}\delta(t-t'-r/c)\exp\bigg[i c(k_0-k_z)(t-t')\bigg]\nonumber\\
&=|\varepsilon_{k_0}|^2 \int_{-R}^{R}dr \frac{1}{2c}\frac{R}{c}|\varepsilon_{k_0}|^2 \exp\bigg[i c(k_0-k_z)r/c\bigg]\nonumber \\
&=|\varepsilon_{k_0}|^2\frac{\sin\left((k_0-k_z)R\right)}{c(k_0-k_z)}
\end{align}

Here $R$ is the radius of the atomic cloud in the cavity which is much larger than the cavity wavelength. 
In the last line we have used that $|\varepsilon_k|^2$ does not vary significantly around $k\sim k_0$ and pull it out of the integral
\begin{align}
\int_0^{\infty} dk |\varepsilon_{k}|^2\exp\bigg[ic(k-k_0)(t-t'-r/c)\bigg]&=\frac{2\pi}{c}\delta(t-t'-r/c)|\varepsilon_{k_0}|^2.
\end{align}
As a result, one obtains
\begin{align}
\int_0^t dt' \sum_{k} |\varepsilon_k|^2 e^{i(\b{k}-\b{k}_0)(\b{r}_{\ell}-\b{r}_{\ell'})}\zeta_k &= \frac{\sin\left((k_0-k_z)R\right)}{c(k_0-k_z)}|\varepsilon_{k_0}|^2 =\frac{2\pi D(k_z)}{2\pi D(k_z)}\frac{\sin\left((k_0-k_z)R\right)}{c(k_0-k_z)}\frac{|\varepsilon_{k_0}|^2}{|\varepsilon_{k_z}|^2}|\varepsilon_{k_z}|^2\nonumber\\
&=\gamma\frac{\sin\left((k_0-k_z)R\right)}{c(k_0-k_z)}\frac{|\varepsilon_{k_0}|^2}{2\pi D(k_z) |\varepsilon_{k_z}|^2}=\gamma\frac{3}{8\pi}\left(\frac{\lambda_z^2}{4\pi R^3}\right)\frac{|\varepsilon_{k_0}|^2}{|\varepsilon_{k_z}|^2}\frac{\sin\left((k_0-k_z)R\right)}{(k_0-k_z)}\equiv \gamma\alpha
\label{Eq:AlphaCollectiveContribution}
\end{align}
here, the volume of the atomic sample is taken to be $V=4/3 \pi R^3$ and the density of states $D(k_z)=Vk_z^2/\pi^2 c$ and $\gamma=2\pi |\varepsilon_{k_z}|^2D(k_z)$. If the difference between the two wavenumbers $k_0$ and $k_z$ is small, one can expand the $\sin$ function to first order to reproduce the result for the strength of the collective decay in \cite{Scully2015} obtained from a wavefunction picture, i.e.\,
\begin{align}
\lim_{k_z \to k_0}\alpha=\frac{3}{8\pi}\left(\frac{\lambda_0^2}{4\pi R^2}\right)
\label{EQ:geometricfactor}
\end{align} 
Eq.\,\eqref{Eq:AlphaCollectiveContribution} and Eq.\,\eqref{EQ:geometricfactor} determine the strength of collective losses in a large sample limit $R\gg \lambda_0$. Using Eq.\,\eqref{Eq:AlphaCollectiveContribution} in Eq.\,\eqref{EQ:NonLocalLindbladContribution} leads to 
\begin{align}
\partial_t \rho_{\rm sys}(t)=&\mathcal{L}_\gamma[\rho]=\gamma\sum_{\ell=1}^N \left(\sigma^{-}_{\ell}\rho_t\sigma^{+}_{\ell}-\frac{1}{2}\{\sigma^{+}_{\ell}\sigma^{-}_{\ell},\rho_t\}\right)+\gamma \alpha \sum_{\ell'\neq\ell}\left(\sigma^{-}_{\ell'}\rho_t\sigma^{+}_{\ell}-\frac{1}{2}\{\sigma^{+}_{\ell'}\sigma^{-}_{\ell},\rho_t\}\right).
\label{EQ:LocalandNonLocalDissipators}
\end{align}
The established decay rates for a single excitation are reproduced by $\mathcal{L}_\gamma$  \cite{Scully2015}, i.e. one finds the single atom decay rate $\gamma$, the decay rate of a superradiant state of $N$ atoms to be $\gamma(1+\alpha(N-1))$ and the decay rate of a subradiant state (see e.g. \cite{Scully2015}) to be $\gamma(1-\alpha)$.
The prefactor $\alpha$ is bounded, $0\le\alpha\le1$, and depends on the cavity geometry as shown in Eq.\,\eqref{EQ:geometricfactor}. 

In order to define a sensible thermodynamic limit $N\rightarrow\infty$, both the average energy and loss rate per particle have to remain finite. The collective loss rate, however, scales as $\alpha N$ and thus we
define $\alpha=\beta/N$ where $\lim_{N\rightarrow\infty}\beta=const.$ has to remain constant in the thermodynamic limit. This leads to the Lindblad superoperator $\mathcal{L}_\gamma$ as given in the main text in Eq.\,(2) and in Eq.\,(3). In an experimental setup, $\beta$ is then determined by the cavity geometry factor $\alpha$ and the experimentally relevant number of atoms $N=N_{\rm exp}$. This is equivalent to the Dicke atom-photon coupling, which is set to scale $\sim g/\sqrt{N}\sum_{\ell=1}^N \sigma^x_{\ell}(a+a^{\dagger})$. \\

From the above derivation it can be seen that the collective atomic loss channel and the individual atomic loss channel are derived from the same Hamiltonian that couples the system degrees of freedom with the electromagnetic vacuum. The collective atomic loss channel does therefore not introduce any new characteristic time scales that would call the time-local Lindblad structure and thus the Born-Markov approximations into question.\\

Below we quote the experimental values for the parameter set $(\omega_0,\omega_z,\kappa,\gamma)=(100(5), 77(2), 100, 0.075-0.3)$ kHz as obtained for a quantum optical realisation of the Dicke model, see \cite{Zhiqiang17} for details. Here $\kappa$ is the half-width-half-maximum linewidth of the cavity, $\gamma$ is the effective rate of spontaneous emission per atom $\omega_0$ is an effective frequency for cavity photons and $\omega_z$ is an effective frequency for the level splitting of the atoms. In this experiment, the typical number of atoms $N$ loaded into the trap was around $N\sim 10^5$. An estimation of the cavity coupling constant $\alpha\sim 6 \cdot 10^{-7}$ from experimental values in \cite{Zhiqiang17} and in \cite{Baden12} leads to a collective loss rate of about $\gamma \beta=\gamma \alpha N \sim (0.0045-0.018)$ kHz. The cooperative loss rate for the current experimental setup is not yet large enough to lead to a competition with the effective atom-atom coupling strength $J$ that is on the order of the atom and cavity frequencies $\gamma \beta \ll J\sim \omega_0$. Future experiments into the dissipative Dicke model might be able to access the regime where $\gamma \beta \sim J$ that is necessary to experimentally measure the bistability regime and the fluctuation induced dynamics.

\twocolumngrid

\bibliography{bib_Dicke1st_2018}

\end{document}